\documentclass[12pt,a4paper]{article}

\usepackage{amsmath}
\usepackage[mathcal]{euscript}
\usepackage{latexsym}
\usepackage{textcomp}
\usepackage{slashed}
\usepackage{tikz}
\usepackage{xspace}
\usepackage{setspace}

\newcommand{\Dd}[0]{\frac{\Delta}{2}}

\begin{document}

\begin{flushright}
IRFU-14-54
\end{flushright}

\begin{center}
\begin{spacing}{1.5}

{\Large\bf Modeling the pion \\Generalized Parton Distribution} \\
\end{spacing}
\vskip 10mm

C.~M\scshape {ezrag}\footnote{cedric.mezrag@cea.fr}
\\[1em]
{\small {\it IRFU/Service de Physique Nucl\'eaire \\ CEA Saclay, F-91191 Gif-sur-Yvette, France}}\\

%\date
\end{center}
\vskip 8mm 
\begin{abstract}
\noindent We compute the pion Generalized Parton Distribution (GPD) in a valence dressed quarks approach. We model the Mellin moments of the GPD using Ans\"{a}tze for Green functions inspired by the numerical solutions of the Dyson-Schwinger Equations (DSE) and the Bethe-Salpeter Equation (BSE). Then, the GPD is reconstructed from its Mellin moment using the Double Distribution (DD) formalism. The agreement with available experimental data is very good.
\vskip 4mm
\end{abstract}

\section{Introduction}	
Introduced in the 1990s \cite{Mueller:1998fv,Ji:1996nm,Radyushkin:1997ki}, GPDs have been intensively studied both theoretically and experimentally. Concerning the proton GPD, several models have emerged \cite{Goloskokov:2005sd,Guidal:2004nd,Kumericki:2009uq,Mezrag:2013mya,Polyakov:2008xm,Goldstein:2010gu} based on different kind of parametrizations, and fitted to data. Here we focus on the pion GPD, which we model in an original way through the triangle approximation, but using propagators and vertices coming from the numerical solutions of the DSE and BSE. This approach have been successful in the case of the pion Parton Distribution Amplitude (PDA)\cite{Chang:2013pq}.
In the first section, the details of the model will be given. Mellin moment will be computed using functional forms coming from the solutions of the DSE and BSE. The results will also be compared with available experimental data. In the second sections, the main point will be the reconstruction of the GPD itself thanks to its Mellin moments. We will details the DD method which reveals itself very convenient. In the third section, symmetry questions will be highlighted allowing us to go beyond the triangle diagram approximation.

\section{Computing the Mellin moments}
In the pion case, the GPD $H$ is formally defined as the Fourier transform of a non local matrix element:
\begin{equation}
   \label{eq:DefinitionGPDH}
   H(x,\xi,t) = \frac{1}{2} \int \frac{\textrm{d}z^-}{2\pi} \, e^{i x P^+ z^-} \left\langle P+\frac{\Delta}{2} \left| \bar{q}\left(-\frac{z}{2}\right)\gamma^+q\left(\frac{z}{2}\right) \right |P-\frac{\Delta}{2}\right\rangle_{z^+=0,z_\perp=0}
\end{equation}

\begin{figure}[h]
  \centering
  \includegraphics[width=0.4\textwidth]{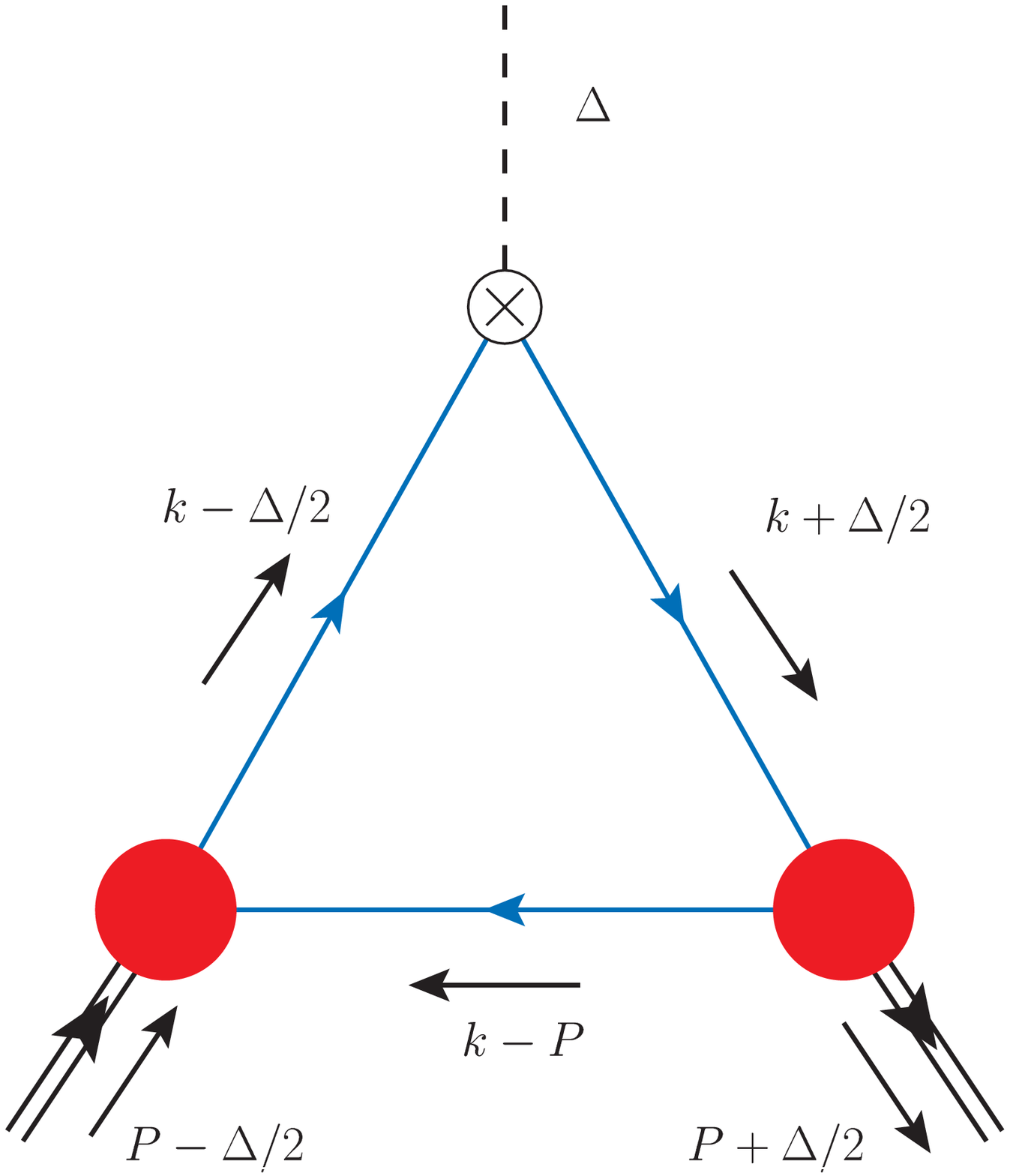}
  \quad
  \includegraphics[width=0.4\textwidth]{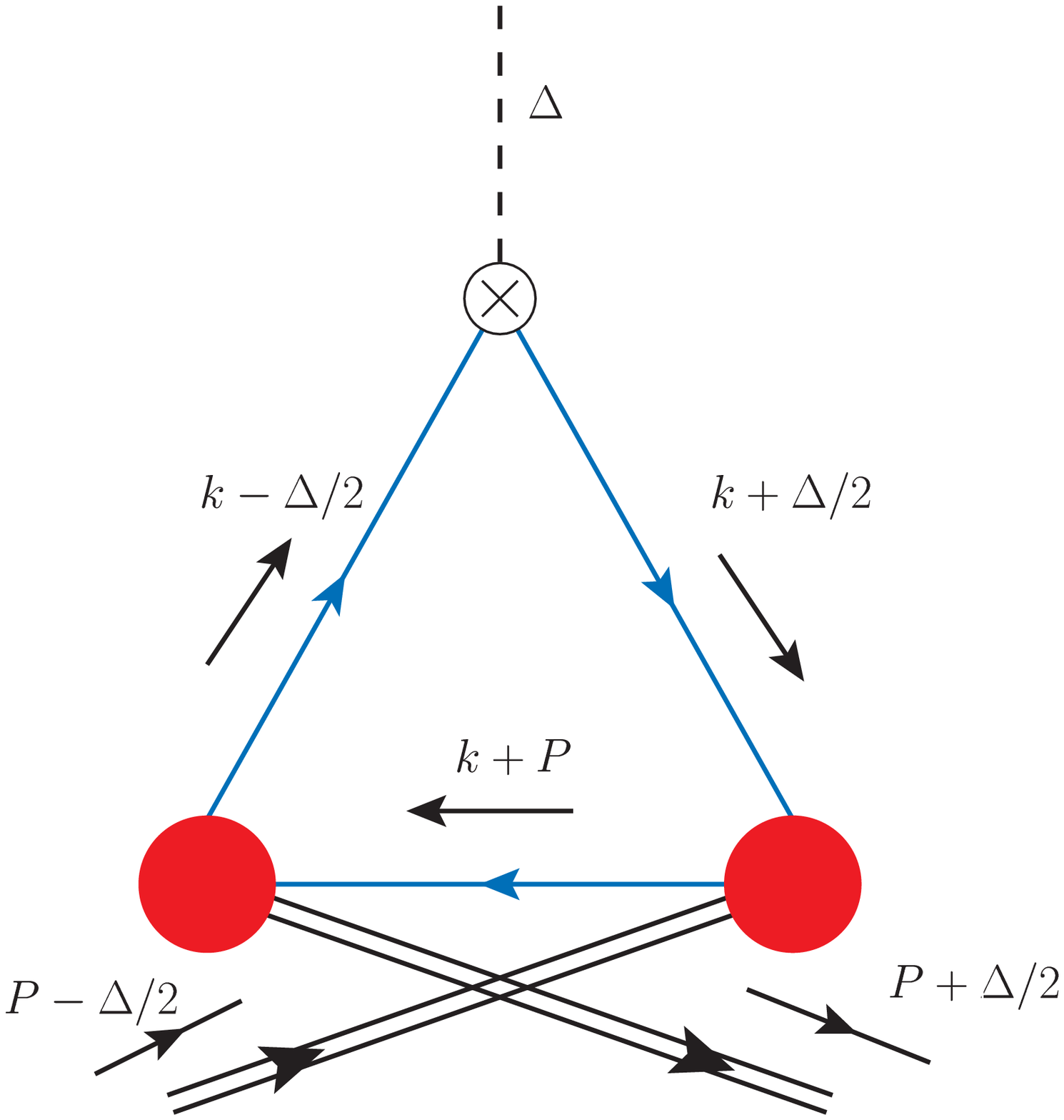}
  \caption{Model of GPD Mellin moment in the triangle approximation for a valence dressed quark.\emph{Left panel}: quark case. \emph{Right Panel}: anti-quark case.}
  \label{fig:TriangleDiagram}
\end{figure}

\begin{figure}[h]
  \centering
  \includegraphics[width=0.4\textwidth]{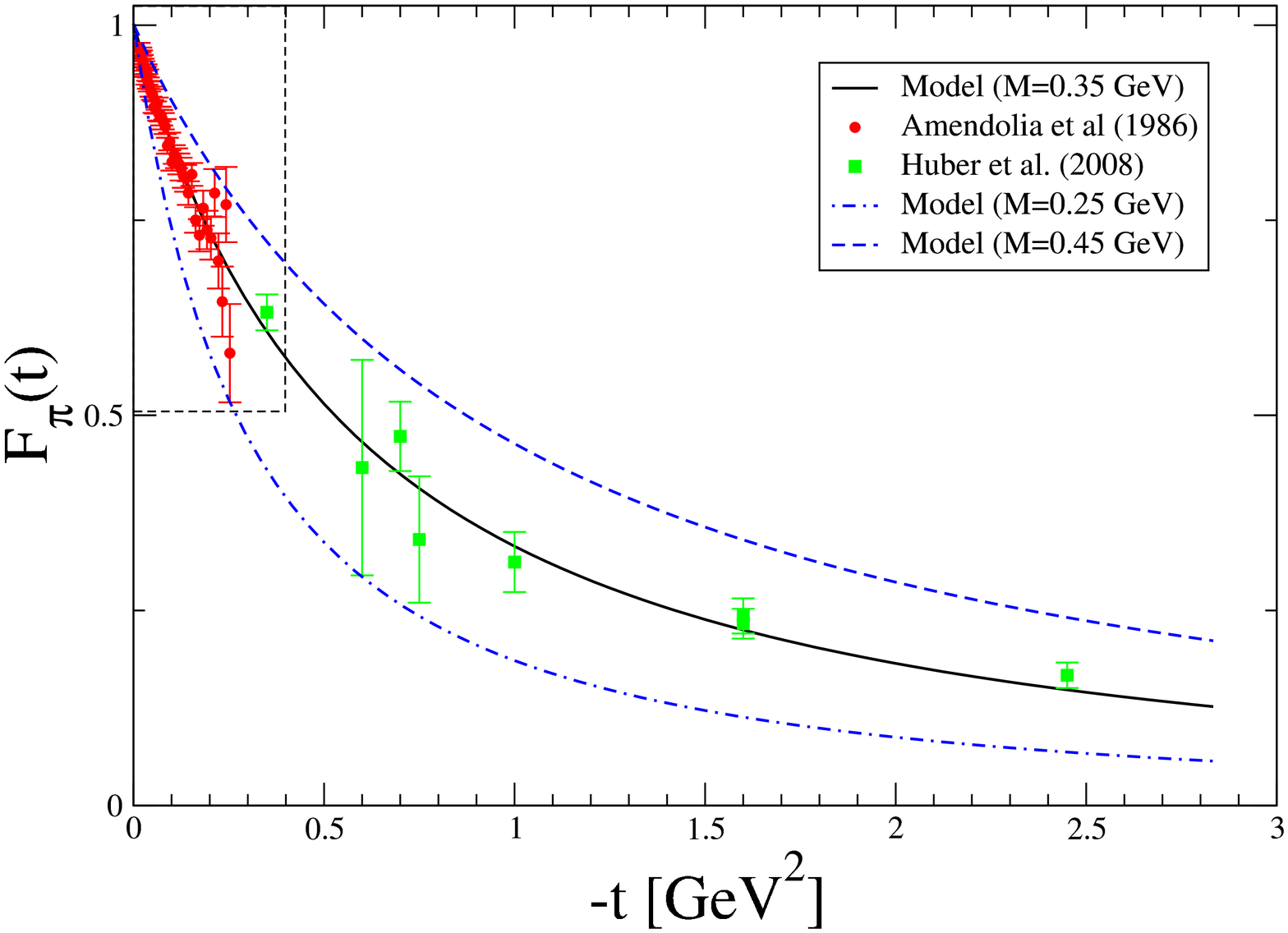}
  \quad
  \includegraphics[width=0.4\textwidth]{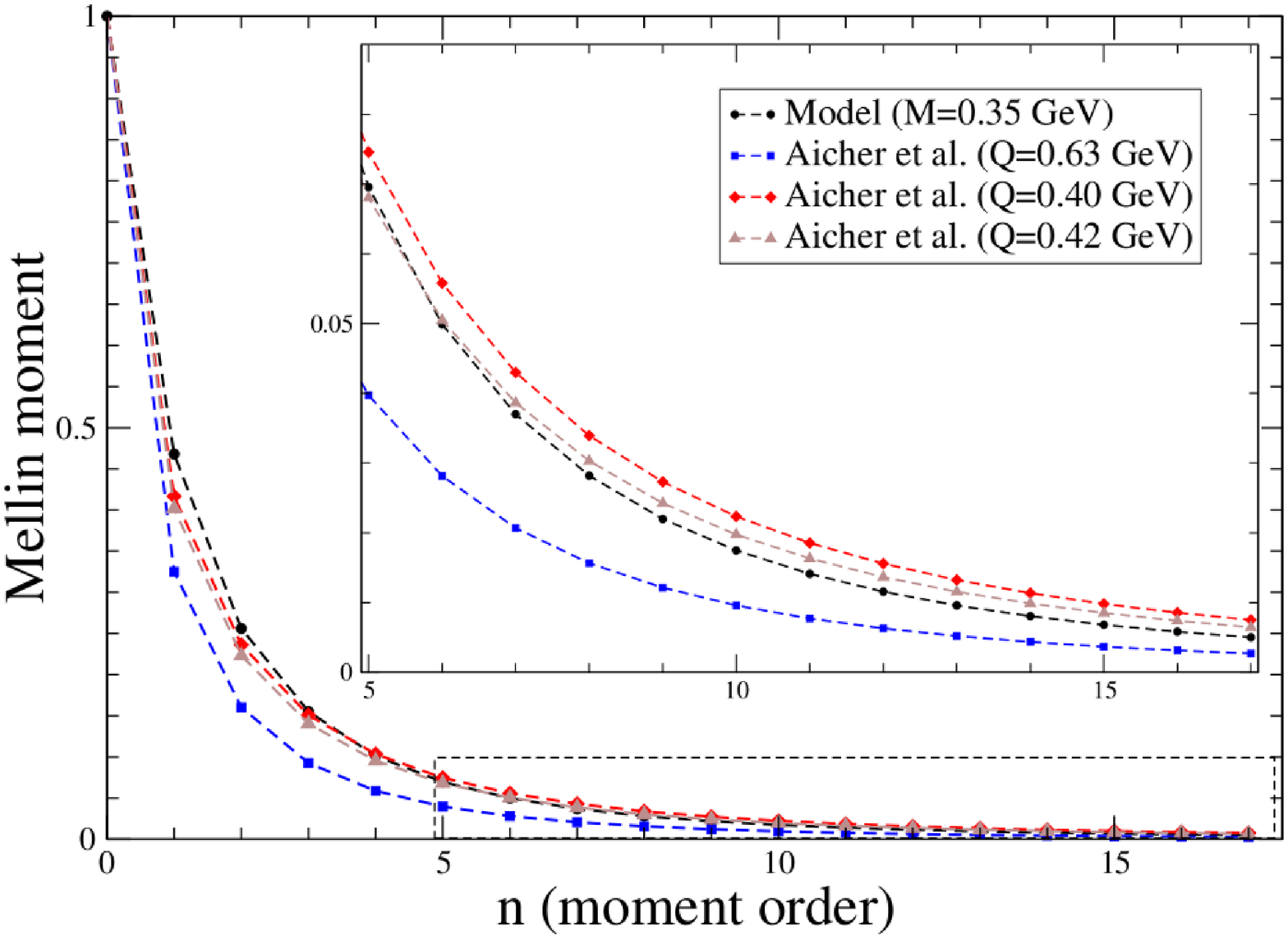}
  \caption{\emph{Left Panel}: Comparison of our model with experimental data for the pion form factor. \emph{Right panel}: Comparison of our model with an PDF fit to data, before and after evolution at low scale.}
  \label{fig:ComparisonExperimentalData}
\end{figure}

Instead of modeling directly the GPD $H(x,\xi,t)$, we focus on the computation of its Mellin moments $\mathcal{M}_n(\xi,t)$ defined as:
\begin{equation}
  \label{eq:DefinitionMellinMoments}
  \mathcal{M}_n(\xi,t) = \int \textrm{d}x~x^nH(x,\xi,t)
\end{equation}
and which are even polynomials in $\xi$ of degree at most $n+1$. In order to model those object, the so-called triangle diagram approximation is used here, as depicted on Fig. \ref{fig:TriangleDiagram}.
Computing this diagram naturally leads to an expression for the Mellin moments as:
 \begin{eqnarray}
  \label{eq:IntegralMellinMoments}
  2( P \cdot n )^{m+1} \mathcal{M}_m(\xi,t) & = & \textrm{Tr} \int \frac{\mathrm{d}^4k}{(2\pi)^4} \, ( k \cdot n )^m i\Gamma_\pi( k -\Dd, P - \Dd )~ S( k - \frac{\Delta}{2} ) \nonumber\\
         & & i\gamma \cdot n~ S( k+\frac{\Delta}{2} )~i\bar{\Gamma}_\pi( k+\Dd,  P+\frac{\Delta}{2} )~ S( k - P ),
\end{eqnarray}
where
\begin{eqnarray}
S( p ) 
& = & \big[ - i \gamma \cdot p + M \big] \Delta_M( p ^2 ), \label{eq:QuarkPropagator} \\
\Delta_M( s )
& = & \frac{1}{s + M^2}, \label{eq:PropagatorMassTerm} \\
\Gamma_\pi( k, p )
& = & i \gamma_5 \frac{M}{f_\pi}M^{2\nu} \int_{-1}^{+1} \mathrm{d}z \, \rho_\nu( z ) \ 
\left[\Delta_M( k_{+z}^2 )\right]^\nu; \label{eq:Vertex} \\
\rho_\nu( z ) & = & R_\nu ( 1 - z^2 )^\nu, \label{eq:RhoFunction} 
\end{eqnarray}
have been inspired by the numerical solutions of the DSE-BSE\cite{Chang:2013pq}. $\Gamma_\pi(k,P)$ is the amputated Bethe-Salpeter pion vertex, $S(k)$ is the quark propagator and $\textrm{Tr}$ stands for the trace on color, flavour and Dirac indices. $M$ is a free parameter which can be seen as the effective quark mass in this algebraic model. $\nu$ is also a free parameter, but in all the following work, it will be fixed to 1, a value which allows one to recover the pion asymptotic DA\cite{Chang:2013pq}. $f_\pi$ is the pion decay constant and $R_\nu$ is a normalization constant. Comparison with experimental data can be done on the form factor \cite{Amendolia:1986wj,Huber:2008id}, which is the Mellin moment of order 0 of the GPD, and the PDF. Unfortunately, no data at $\xi \neq 0$ is available now, despite some pioneering work \cite{Amrath:2008vx}. The agreement with experimental data shown on Fig. \ref{fig:ComparisonExperimentalData} is very encouraging for this algebraic model \cite{Mezrag:2014tva}.

\section{Double Distributions}
\begin{figure}[h]
  \centering
  \includegraphics[width=0.4\textwidth]{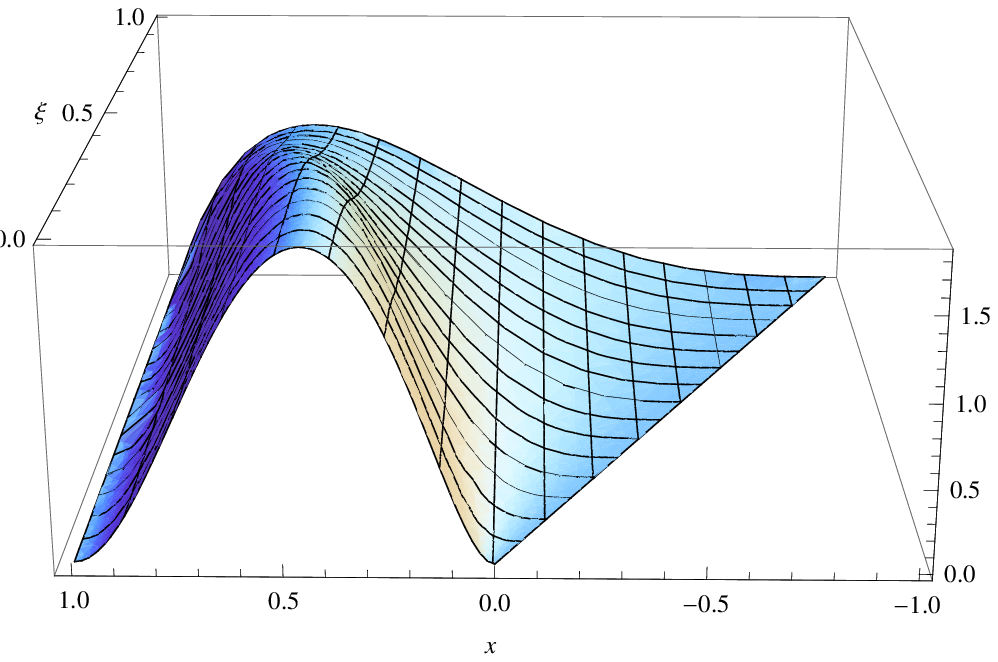}
  \quad
  \includegraphics[width=0.4\textwidth]{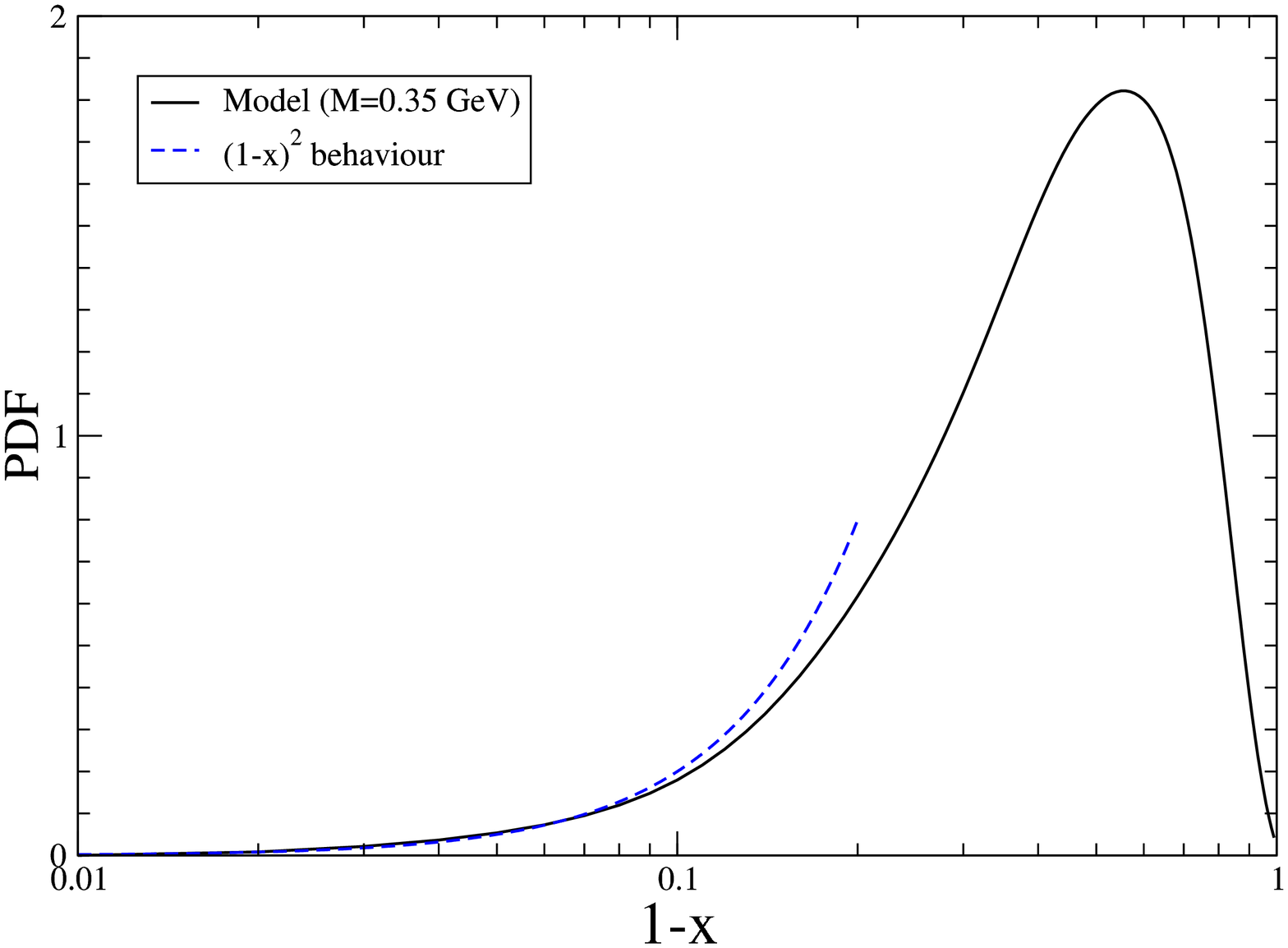}
  \caption{\emph{Left Panel} Full GPD reconstruction at vanishing t.This plot illustrates the properties of support and continuity. \emph{Right Panel}: Behaviour of the PDF at large $x$. We get back here the pertubative behaviour.}
  \label{fig:GPD}
\end{figure}

DDs have been introduced very early in order to model GPDs\cite{Mueller:1998fv,Radyushkin:1998es,Radyushkin:1998bz} because they ensure the polynomiality property of the Mellin moments. We remind the reader that the link between the GPD $H$ and the DD is a Radon transform:
\begin{equation}
  \label{eq:DefinitionDD}
  H(x,\xi,t) = \int_{|\alpha|+|\beta|\le1}\textrm{d}\beta\textrm{d}\alpha~\delta(x-\beta-\alpha \xi)\left(F(\beta,\alpha,t)+\xi G(\beta;\alpha,t)\right)
\end{equation}
It is possible within our model, to rewrite the Mellin moments of the GPD in terms of Mellin moments of the DDs. From there, one can identify for any order of the Mellin moments the DDs, and thus get back the GPD through the Radon transform \cite{Mezrag:2014tva}. Consequently, one can get an analytic expression for the GPD, and thus check the fundamental properties as polynomiality of the Mellin moment, continuity at $x=\xi$ and the support property. All of them are fulfilled in our approach as it can be noticed on Fig. \ref{fig:GPD}

\section{Symmetry considerations}

\subsection{PDF case}

\begin{figure}[h]
  \centering
  \includegraphics[width=0.4\textwidth]{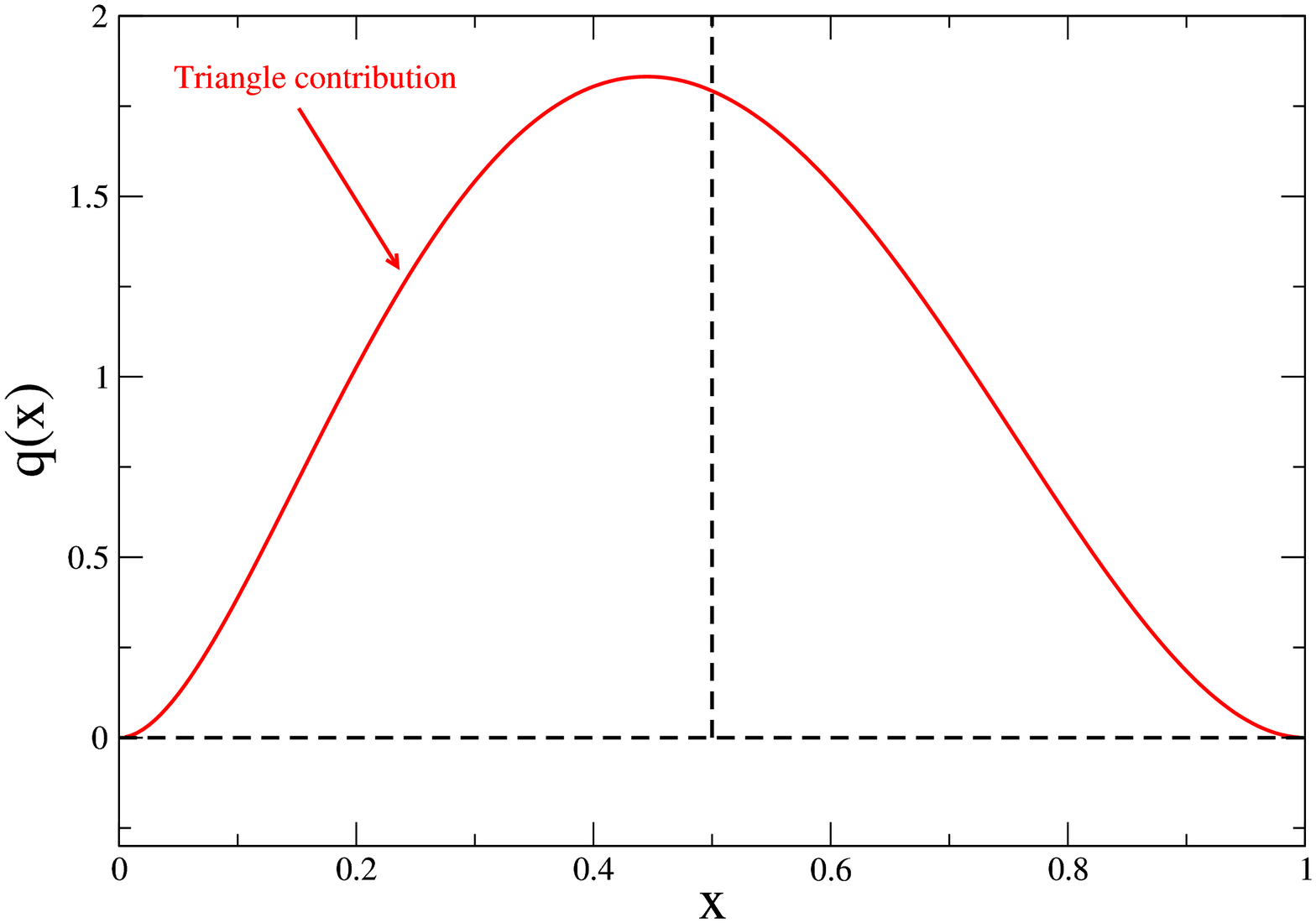}
  \quad
  \includegraphics[width=0.4\textwidth]{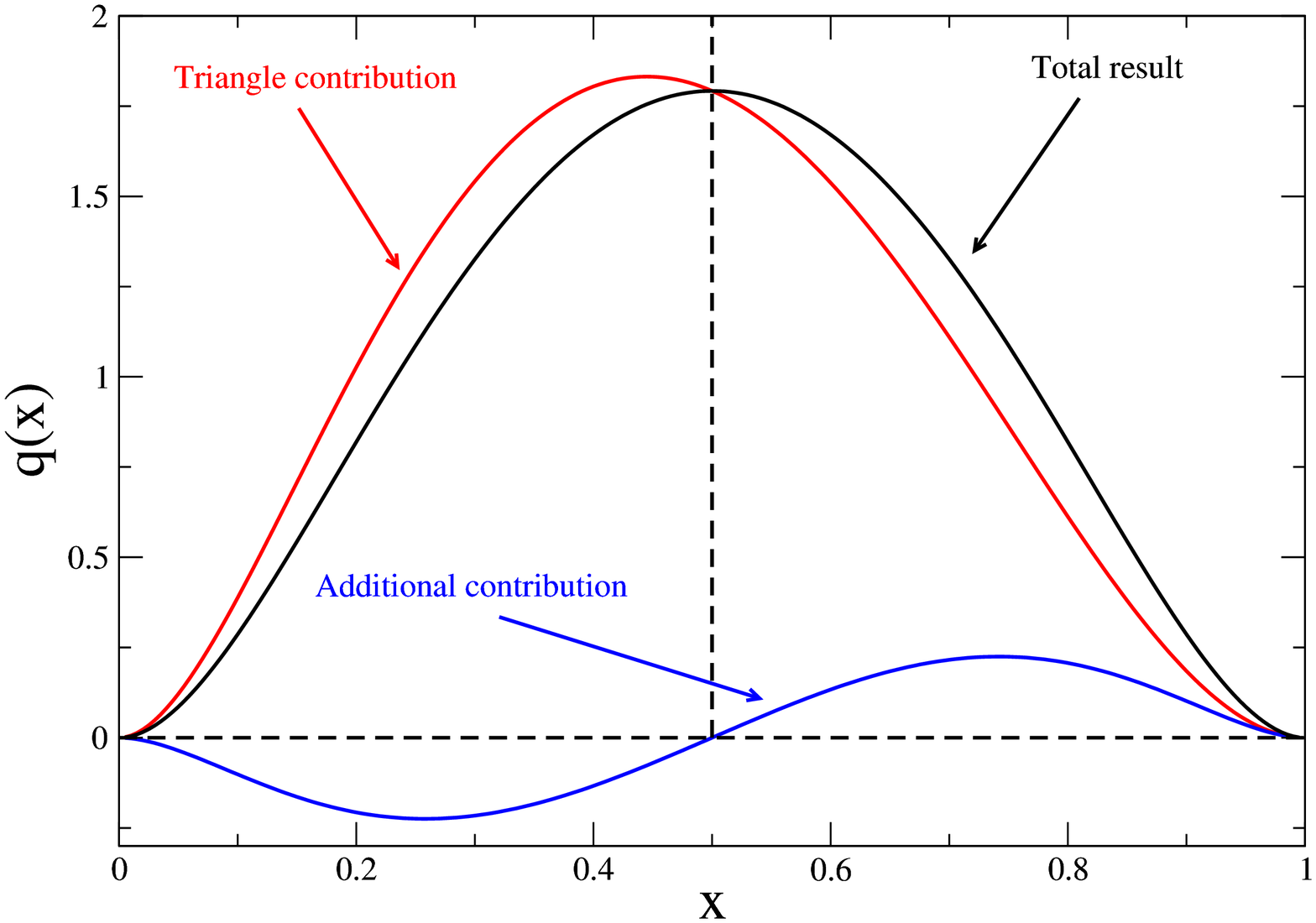}
  \caption{\emph{Left Panel}: PDF obtained with the triangle diagram only. The symmetry with respect to $x=\frac{1}{2}$ is slightly broken. \emph{Right Panel}: Additional contribution beyond the triangle diagram. The sum of the two contributions is symmetric}
  \label{fig:SymmetryPDF}
\end{figure}

Getting an analytic expression for the GPD allows also one to check the consistency of the different assumptions. One of them is the fact that due to a combination of isospin symmetry, charge symmetry and the two-body nature of our problem (two dressed quarks), the PDF must be symmetric with respect to $x=\frac{1}{2}$. As it can be seen on left panel of Fig. \ref{fig:SymmetryPDF}, this symmety property is slightly broken. 
This due to the assumption to work with a triangle diagram\cite{Chang:2014lva}. It is also possible to take into account the twist two operators directly when computing the Bethe-Salpeter amputated vertex. Adding such a contribution gives back a symmetric PDF, as shown on Fig. \ref{fig:SymmetryPDF}

\subsection{Soft pion theorem}

There is an interesting link between the pion GPD and the pion PDA \cite{Polyakov:1998ze}. Denoting the pion PDA by $\phi(z)$, one can show that:
\begin{equation}
  \label{eq:SoftPionTheorem}
  H(x,\xi=1,t=0) = \frac{1}{2} \phi \left(\frac{1+x}{2}\right).
\end{equation}
Our model failed to reproduce this property. There is no mystery behind that. Indeed, the triangle diagram is consistent enough with the rainbow ladder truncation of the DSE, such that it will give automatically the soft pion theorem, provided that the Axial-Vector Ward Takahashi Identities (AVWTI) are fulfilled \cite{Mezrag:2014jka}. In this algebraic model, it is not the case. But it is enough to upgrade the model by using directly the numerical solutions of the BSE-DSE to fulfill the AVWTI, and thus the soft pion theorem.

\section{Conclusions and outlook}
Our study has shown that it is indeed possible to build a new model of GPD based on the Dyson-Schwinger equations which will respects most of the theoretical constraints. It has been shown that, it was possible to pursue all the computations analytically providing that one uses the DD formalism. Furthermore the very satisfactory comparison with experimental data is a key encouraging point to go beyond this algebraic model and build another model using the numerical solutions of the DSE and BSE. It also pave the way for a proton GPD, based on a quark-diquark model.

\section*{Acknowledgments}

I thank my collaborators L. Chang, H. Moutarde, C.D. Roberts, J. Rodriguez-Quintero, F. Sabati\'e and P.C. Tandy with whom this work has been done. I also thank A. Besse, P. Fromholz, P. Kroll, J-Ph. Landsberg, C. Lorc\'e, B. Pire and S. Wallon for valuable discussions.  
This work is partly supported by the Commissariat \`a l'Energie Atomique, the Joint Research Activity "Study of Strongly Interacting Matter" (acronym HadronPhysics3, Grant Agreement n.283286) under the Seventh Framework Programme of the European Community, by the GDR 3034 PH-QCD "Chromodynamique Quantique et Physique des Hadrons", the ANR-12-MONU-0008-01 "PARTONS".

\appendix

\bibliographystyle{unsrt}
\bibliography{/home/cedric/Work/paper/Bibliography}

\end{document}